\documentclass[preprint,12pt]{revtex4}

\usepackage[brazil, english]{babel}
\usepackage[utf8]{inputenc}
\usepackage{graphicx}
\usepackage{epsfig}
\usepackage{amssymb}
\usepackage{amsmath} 
\usepackage{slashed}
\usepackage{natbib}
\usepackage[unicode=true,bookmarks=false,breaklinks=false,pdfborder={0 0 1},colorlinks=true]
 {hyperref}
\hypersetup{
 citecolor=blue,linkcolor=blue,urlcolor=blue}

\graphicspath{%
    {converted_graphics/}
    {/}
}
\begin{document}

\title{Proton and neutron form factors \\
from deformed gravity/gauge duality} 
\author{Miguel Angel Martín Contreras$^1$}
\email{miguelangel.martin@uv.cl}
\author{Eduardo Folco Capossoli$^2$}
\email{eduardo\_capossoli@cp2.g12.br}
\author{Danning Li$^{3}$}
\email{lidanning@jnu.edu.cn} 
\author{Alfredo Vega$^1$}
\email{alfredo.vega@uv.cl} 
\author{Henrique Boschi-Filho$^{4}$}
\email{boschi@if.ufrj.br}  
\affiliation{$^1$Instituto de Física y Astronomía, Universidad de Valpara\'iso, A. Gran Breta\~na 1111, Valpara\'iso, Chile\\
$^2$Departamento de F\'\i sica / Mestrado Profissional em Práticas de Educação Básica (MPPEB), Col\'egio Pedro II, 20.921-903 - Rio de Janeiro-RJ - Brazil\\
$^3$Department of Physics and Siyuan Laboratory, Jinan University, Guangzhou 510632, China\\
 $^4$Instituto de F\'\i sica, Universidade Federal do Rio de Janeiro, 21.941-972 - Rio de Janeiro-RJ - Brazil}

\begin{abstract}
In this work we study the electric and magnetic Sachs form factors for proton and neutron by using a deformed gravity/gauge model. We describe holographically  baryons as well as gauge bosons which are ruled by the parameters $k_{B}$ and $k_\gamma$, associated with   confinement and kinematical energy scales, respectively. Then, we construct the interaction action and calculate the electric and magnetic form factors for the proton and the neutron, and their electromagnetic  sizes. The obtained results are compatible with those found in the literature from experimental and theoretical data.
\end{abstract}


\maketitle


\section{Introduction}

Protons and neutrons are the basic constituents of ordinary nuclear matter and since its discovery in the last century they still attracting much attention from the physics community. Historically located in the middle of the 20th century, many high energy experiments started to scatter leptons (electrons or muons) off  proton or neutron targets probing  hadronic internal structure. These scatterings could be elastic or inelastic. The deep inelastic scattering (DIS) is very important since it revealed that protons are made of quarks. On the other hand, through elastic scattering  scientists could perform experimental measurements related to the form factors of proton and neutron. The electric ($G^E$) and magnetic ($G^M$) form factors  (FF) retain information related to the spatial distribution of the electric and magnetic charges inside hadrons, and then, due to this, these FF became an important observable for better understanding the structure of nuclear matter. The pioneer works \cite{Hofstadter:1953zjy, Hofstadter:1955ae, Yearian:1958shh} produced the first experimental data for both proton and neutron FF. For  reviews on related experimental research, see for instance Refs. \cite{Kelly:2004hm, Perdrisat:2006hj, Arrington:2007ux,  Arrington:2011kb,Bernauer:2013tpr}.
The canonical way to study FF resorts to QCD as can be seen, for instance, in Refs. \cite{Carvalho:2005mu, Arrington:2006zm, Pacetti:2015iqa, Ye:2017gyb}. Unfortunately QCD at the low energy regime cannot be treated perturbatively  since in this case the corresponding coupling constant g$_{YM}>1$. 

 Here, in this work, we choose to study FF using a holographic approach based on the AdS/CFT correspondence \cite{Aharony:1999ti}. This correspondence or duality maps a superconformal and strongly coupled theory in a four-dimensional flat spacetime onto a weak coupled one in a five-dimensional curved spacetime.
  In order to reproduce some features of $SU(3)$-like theories one needs firstly to break the conformal symmetry, and then, to compute, $G^E(q^2)$ and $G^M(q^2)$ for both proton and neutron as functions of the transferred  four-momentum squared $q^2$. In particular, our model, takes into account an exponential factor inserted in the metric of the five-dimensional anti-de Sitter (AdS$_5$) space.

 Although the introduction of this factor into the metric looks like similar to the original softwall model  \cite{Karch:2006pv} (where this factor is introduced into the action) these formulations are different. In this  deformed AdS space formulation one has an effective break of the conformal symmetry for fermions in contrast to the original softwall model. This means that in the original softwall model the dilaton field alone does not generate a discrete fermionic spectrum. The present model is inspired by Refs. \cite{Andreev:2006vy, Andreev:2006ct}
where this warp factor was introduced in the AdS metric to obtain the quark-antiquark potential. Within the holographic context, covering many topics and purposes, several authors have used some kind of deformation in the AdS space metric as can be seen in Refs. \cite{Ghoroku:2005vt, Afonin:2012jn, Rinaldi:2017wdn, Bruni:2018dqm, Afonin:2018era,  FolcoCapossoli:2019imm, Rinaldi:2020ssz, Caldeira:2020sot, FolcoCapossoli:2020pks, Caldeira:2020rir, MartinContreras:2021yfz,  Caldeira:2021izy,  Afonin:2021cwo}.

 As the AdS/CFT correspondence itself has proved a successful alternative way to deal with QCD at low energy, various authors have used it to study nucleon FF  by using different frameworks as can seen through an incomplete list in Refs. \cite{ Abidin:2009hr, Abidin:2009aj, Bayona:2011xj, Chakrabarti:2013dda,  Gutsche:2012bp, Gutsche:2014yea, Gutsche:2015xva, Mamedov:2016ype, Vega:2018lym, Mamo:2021jhj}.

This work is organized as follows: in section \ref{esff} we review briefly the main properties of elastic scattering and form factors.  In section \ref{defholo} we present our deformed AdS space model and a  holographic descriptions of baryons and gauge bosons. In section \ref{ff} we compute the interaction between   vector and spinor fields. From this interaction action we achieve the electric and magnetic Sachs form factors for the proton and the neutron, and compute the   electromagnetic  size of these nucleons. In section \ref{results} we present our discussions and comments on the computed form factors and the nucleons size.

\section{The elastic scattering and the electromagnetic form factors}\label{esff}

In this section we present a brief review on elastic scattering processes which relates a lepton $\ell$ (electron or muon) scattered off a target hadron (proton or neutron) by the exchanging a virtual photon with four-momentum $q^\mu$. As these are elastic processes, the initial and final hadronic states are the same. For a review one can see, for instance, Ref. \cite{Pacetti:2015iqa}.
%
%
%
Considering the matrix element for elastic electron-proton scattering, up to some constants factors, as:
\begin{equation}\label{amp}
i\,\mathcal{M}_{e^{-}p\to e^{-}p} \sim \left[ {\bar u}(k') i \gamma^{\mu} u(k)\right] \left(\frac{- i e^2 g_{\mu \nu}}{q^2}\right)\left[ {\bar u}(P') i \Gamma^{\nu} u(P)\right] \sim \frac{e^2}{q^2}J^{\mu}_e J_{\mu p}\,, 
\end{equation}
\noindent where $k, k'$ and $P, P'$ are the initial and final momentum for the electron and the proton, respectively, and $J^{\mu}_e$ is the electronic  electromagnetic current.  
The hadronic electromagnetic current can be expressed in terms of the Dirac $(F_1)$ and Pauli ($F_2$) form factors, as
\begin{equation}\label{form-factor-matrix-1}
    J_{\mu p} = e \,{\bar u}(P') \left[ \gamma^{\mu} F_1(q^2) + \frac{1}{2 m_p} F_2(q^2)\sigma^{\mu \nu} q_{\nu}\right]u(P)\,.
\end{equation}

By using the Sachs notation \cite{Sachs:1962zzc} one can write the electric and magnetic form factors, 
\begin{equation}
    G^E(q^2) = F_1(q^2) + \frac{q^2}{4 m_P} F_2(q^2), \,\,\,G^M(q^2) = F_1(q^2) +
    \, F_2(q^2)\,,
\end{equation}
 respectively.


\section{Deformed holographic description of hadrons and electromagnetic current}\label{defholo}

In this section we will present our deformed AdS/QCD model which will be used to describe the proton and neutron as well as to calculate the electromagnetic form factors, $G^E(q^2)$ and $G^M(q^2)$. For our purposes, the five-dimensional action for the fields can be written as: 
\begin{equation}\label{acao_soft}
S = \int d^{5} x \sqrt{-g^I} \; {\cal L}
\end{equation}

\noindent where ${\cal L}$ is the Lagrangean density, and $g^I$ is the determinant of the metric $g^I_{mn}$ of the deformed five-dimensional AdS space, written as:
\begin{equation}\label{gs}
ds^2 = g^I_{mn} dx^m dx^n= e^{2A_I(z)} \, (dz^2 + \eta_{\mu \nu}dy^\mu dy^\nu)\,, 
\end{equation}

\noindent where $z$ is the holographic coordinate and $y^\mu$ describe the four-dimensional Minkowski spacetime, which metric 
  $\eta_{\mu \nu}$ has signature $(-,+,+,+)$. The indices $m, n, \cdots$ represent the five-dimensional space, split  into $\mu, \nu, \cdots$ for the Minkowski spacetime and the holographic $z$ coordinate. We  write the warp factor as:
\begin{equation}
    A_I(z) = -\log z + \frac{k_I}{2}\, z^2\,,  \label{A}
\end{equation}
\noindent where the constant $k_I$, with index $I = B, \gamma$, is associated with baryons and the photon, respectively. It has dimension of mass squared and in the case of baryons it is related to a QCD mass scale while for photons it is a kinematical scale. 
Note that the  metric given by Eqs.  \eqref{gs} and \eqref{A}  represents a deformed AdS space because of the presence of the exponential warp factor $e^{k_I z^2}$, and the case $k_I=0$ would be a pure $AdS_5$ space. 

Here it is worthwhile to make a brief discussion on two  parameters, $k_B$ and $k_{\gamma}$. As discussed in Ref. \cite{MartinContreras:2021yfz} the nucleons and the photon do not experience the same geometry which is encoded in these parameters. The first one, $k_B$ is fixed to get both masses for proton and neutron. Besides it is related to the confining IR scale. The second one, $k_{\gamma}$ is related to the virtual photon, and usually defines an energy scale of the scattering process as can be seen, for instance,  in Refs.  \cite{Polchinski:2002jw, Braga:2011wa, Capossoli:2015sfa, FolcoCapossoli:2020pks}.

Baryonic states in the deformed AdS space will be described by the action:  
\begin{equation}\label{diracfield}
S =  \int d^{5} x\sqrt{g^B} \; \bar{\Psi}({\slashed D} - m_5 ) \Psi\,, 
\end{equation}
where the operator ${\slashed D} \equiv g^{mn}_B e^{a}_n \gamma_a \left( \partial_{m} + \frac{1}{2} \omega^{bc}_{m}\, \Sigma_{bc} \right)$
\noindent with $\gamma_a = (\gamma_{\mu}, \gamma_5)$, $\left\lbrace \gamma_a, \gamma_b \right\rbrace = 2 \eta_{ab} $, and $\Sigma_{\mu 5} = \frac{1}{4} \left[ \gamma_{\mu}, \gamma_5\right]$. 
The Dirac's gamma matrices are represented by $\gamma_\mu$ and we will use use $a, b, c$ to represent indexes in flat space,  $m, n, p, q$ to represent  indexes in the deformed $AdS_5$ space, and $\mu, \nu$ to represent the Minkowski space.
Thus, the vielbein are given by $e^{a}_m = e^{A_B(z)}\, \delta^{a}_m$, $e^{m}_a = e^{-A_B(z)} \,\delta^m_{a}$ and $e^{ma} = e^{-A_B(z)}\, \eta^{ma}$, with $m = 0, 1, 2, 3, 5$. For the non-vanishing spin connection $\omega^{\mu \nu}_{m}$ components in deformed AdS$_5$, one has $\omega^{5 \nu}_{\mu} = - \omega^{\nu 5} _{\mu} = \partial_z A_B(z) \delta^{\nu}_{\mu}$. Besides the non-vanishing Christoffel symbols used in this work are $\Gamma_{\mu \nu}^5 = A_B'(z) \eta_{\mu \nu}, \; \; \Gamma_{5 5}^5 = A_B'(z)  \; \; {\rm and} \; \; \Gamma_{\nu 5}^{\mu} = -A_B'(z) \delta^{\mu}_{\nu}.$
%
%
%
%
%
%
%
%
%
%
%
%
These definitions also follow from the pure AdS space, where $A(z) = - \log(z)$, as for instance, in Ref. \cite{Mueck:1998iz}.  
From the action Eq. \eqref{diracfield} one can derive the bulk EOM: 
%
%
%
\begin{equation}\label{newdirac}
\left( e^{-A_B(z)} \gamma^5 \partial_5 + e^{-A_B(z)} \gamma^{\mu}  \partial_{\mu} + 2 A_B'(z)\gamma^5 - m_5\right) \Psi = 0, 
\end{equation}
\noindent where $\partial_5 \equiv \partial_z$, and $m_5$ is the  mass of the bulk state $\Psi$ associated with the boundary baryon. Assuming that the spinor $\Psi$ can be decomposed into right- and left-handed chiral components, one has:
\begin{equation}\label{psi}
\Psi(x^{\mu}, z) = \left[ \frac{1 - \gamma^5}{2} f_L(z) + \frac{1 + \gamma^5}{2} f_R(z)\right] \Psi_{(4)}(x)\,, 
\end{equation}
\noindent where $\Psi_{(4)}(x)$ satisfies the usual Dirac equation $({\slashed \partial} - M)\Psi_{(4)}(x) = 0$ on the flat four-dimensional boundary space. For the left and right modes, one has $\gamma^5 f_{L/R} = \mp f_{L/R}$ and $\gamma^{\mu}  \partial_{\mu} f_R = M f_L$, and $M$ is the four-dimensional fermionic mass. 

Considering that the Kaluza-Klein modes are dual to the chirality
spinors one can expand $\Psi_{L/R} (x^{\mu}, z) = \sum_n f_{L/R}^n  (x^{\mu}) \chi_{L/R}^n (z).$
By using such an expansion with \eqref{psi} in \eqref{newdirac} one gets the coupled equations:
\begin{equation}\label{mix1}
\left(\partial_z + 2 A_B'(z)\, e^{A_B(z)} \pm m_5\,e^{A_B(z)} \right) \chi_{L/R}^n (z) = \pm M_n \chi_{R/L}^n (z)\,. 
\end{equation}
%
%
%

Performing a Bogoliubov transformation $\chi^n_{L/R}(z) = \psi^n(z) e^{-2\, A_B(z)}$ and decoupling Eqs. \eqref{mix1} one gets a Schr\"odinger equation written for both right and left sectors, given by:

%
%
%
\begin{eqnarray}\label{scr}
-\psi_{R/L}''(z) + \left[ m_5^2 e^{2 A_B(z)} \pm m_5 e^{A_B(z)}A_B'(z) \right]\psi_{R/L}(z)  = M_n^2 \psi_{R/L}^n (z),
\end{eqnarray}
\noindent where $M_n$ is the four-dimensional baryon mass for each mode $\psi^n_{R/L}$ and the corresponding potentials are given by:
\begin{equation}\label{potscr}
 V_{R/L}(z) = m_5^2 e^{2 A_B(z)} \pm m_5 e^{A(z)}A_B'(z). 
\end{equation}
One should note that this equation can be applied to any warp factor $A(z)$. The pure AdS space is recovered if one uses $A(z) = - \log(z)$, which leads to analytical solutions. In our case, with $A_B(z)= - \log z + k_Bz^2/2$, Eq. \eqref{A}, we need to resort to numerical methods. 

Before we start the numerical calculation let us recall the AdS/CFT dictionary from which we can relate the conformal dimension $\Delta$ of a boundary operator ${\cal O}$  creating nucleons  with the baryon bulk mass as  $|m_5|= \Delta - 2.$ As we are going to consider an elastic scattering, the initial state is a proton or a neutron, and they will be considered as a single particle 
modeled by the fermion bulk field. 
In this sense, we get $\Delta = 9/2$, which is the usual baryonic dimension. Hence, in this work we have $m_5 = |m_5|= \Delta - 2 = 5/2$.

From the solutions of Eq. \eqref{scr} one can read the initial and final spinor states,  $\Psi_i$ and $\Psi_f$, as linear combinations of the chiral solutions $\psi_{R/L}$, as follows:
\begin{eqnarray}
\Psi_i&=&e^{i\,P\cdot\,y}\,z^2\,e^{-k_B\,z^2}\left[\left(\frac{1+\gamma_5}{2}\right)\psi^i_L(z)+\left(\frac{1-\gamma_5}{2}\right)\,\psi_R^i(z)\right]\,u_{s_i}(P)   \label{psii} \\
\Psi_f&=&e^{i\,P_f\cdot\,y}\,z^2\,e^{-k_B\,z^2}\left[\left(\frac{1+\gamma_5}{2}\right)\psi^f_L(z)+\left(\frac{1-\gamma_5}{2}\right)\,\psi^f_R(z)\right]\,u_{s_f}(P_f),  \label{psif}
\end{eqnarray}
where $s_i$ and $s_f$ are the spin of the initial and final states, respectively. 
%
%

As elastic scattering involves an electromagnetic interaction, at this point we will describe the photon in the deformed AdS space. Firstly, let us to introduce the action for a five-dimensional massless gauge field $\phi^n$ given by:
\begin{equation}\label{f15}
S = -  \int d^{5} x \sqrt{-g_\gamma} \; \frac{1}{4} F^{mn} F_{mn}\,,
\end{equation}
\noindent where $F^{mn} = \partial^m \phi^n - \partial^n \phi^m$. This action leads to the following equations of motion $\partial_m [ \sqrt{-g_\gamma}\; F^{mn}] = 0$. 
By using the gauge fixing $\partial_\mu\,\phi^\mu+e^{-A_{\gamma}}\partial_z\left(e^{A_{\gamma}}\,\phi_z\right)=0,$
where $A_{\gamma}=A_{\gamma}(z)$ is given by Eq. \eqref{A}, with $k_I = k_{\gamma}$ and prime denotes derivative with respect to $z$, one has 
\begin{eqnarray} \label{solem}
\Box\,\phi_\mu+A_{\gamma}'\,\partial_z\,\phi_\mu+\partial_z^2\,\phi_\mu=0\\
\Box\,\phi_z-\partial_z\left(\partial_\mu\,\phi^\mu\right)=0\,. \label{solem2}
\end{eqnarray}
Without loss of generality, we will consider a photon with a particular polarization such that $\eta_ \mu \, q^ \mu =0$. In this sense only  the electromagnetic field component $\phi^\mu$ will contribute to the scattering process as discussed, {\sl e. g.}, in Refs. \cite{Polchinski:2002jw, Braga:2011wa, Capossoli:2015sfa}.
The general solution to equation \eqref{solem} has the following form: 
\begin{eqnarray}
\phi_\mu(z,q)=C^1_{\mu}(y)\, G_{1,2}^{2,0}\left(\frac{k_{\gamma}\,z^2}{2}\left|
\begin{array}{c}
 \frac{q^2}{2 k_{\gamma}}+1 \\
 0,1 \\
\end{array}
\right.\right)-\frac{1}{2} C^2_{\mu}(y)\, k_{\gamma}\, z^2 \, _1F_1\left(1-\frac{q^2}{2 k_{\gamma}};\,2;\,-\frac{k_{\gamma}\,z^2}{2}\right)\,,
\end{eqnarray}
\noindent where $G_{p,q}^{m,n}\left(z\left|
\begin{array}{c}
 a_1 \cdots a_p \\
 b_1 \cdots b_q \\
\end{array}
\right.\right)$ and $_1F_1 (a; b; z)$ are the the Meijer G  and the Kummer confluent hypergeometric functions, respectively. By imposing the boundary condition $\left.\phi_{\mu}(z,y)\right|_{z=0} = \eta_{\mu} e^{i q\cdot y}$, and considering normalizable (square integrable) solutions  that implies $C_\mu^1(y)=0$, one can write:
\begin{eqnarray}\label{phimunorm}
\phi_\mu(z,q)&\equiv& -\frac{\eta_{\mu} e^{i q\cdot y}}{2}\, B(z,q)\,, 
\end{eqnarray}
\noindent where $B(z,q) = k_{\gamma}\, z^2 \, \Gamma{\left[1 - \frac{q^2}{2k_{\gamma}}\right]}\; {\cal U} \left(1-\frac{q^2}{2 k_{\gamma}};\,2;\,-\frac{k_{\gamma}\,z^2}{2}\right) $,  $\Gamma[a]$ and ${\cal U} (a, b,z)$ are the Gamma and Tricomi hypergeometric functions, respectively. Notice that the above equation represents the solution for the electromagnetic field that will be used to compute interaction action in the next section.

\section{Nucleon form factors in the deformed background}\label{ff}

Here in this section, we proceed to compute the electric and magnetic proton and neutron form factors and their corresponding electric and magnetic  sizes within our deformed AdS$_5$ background. 

Holographically speaking, form factors are extracted from interaction terms written in the bulk action.  These interaction terms are defined on the Fock states of baryonic on-shell action, summarized in Eqs.  \eqref{psii} and \eqref{psif}. The electromagnetic interaction is represented by the U(1) massless field in the bulk, whose solution was depicted in Eq.  \eqref{phimunorm}.  At the boundary,  for spin 1/2 hadrons, the electromagnetic form factors $F_1(q)$ and $F_2(q)$ come from the electromagnetic current matrix elements described in Eq. \eqref{form-factor-matrix-1}.

In order to compute nucleon form factors, we assume that the holographic interaction term has the following structure
\begin{equation}\label{sint}
I_\text{int}=\int{d^5x\,\sqrt{-g^B}\,\left\{\bar{\psi}_f\,\Gamma^m\,\phi_m\,\psi_i+\frac{i\,\eta_{N}}{2}\bar{\psi}_f\,\left[\Gamma^m,\Gamma^n\right]\,F_{mn}\,\psi_i\right\}},    
\end{equation}
where $\eta_{N}$ is a parameter that labels the nucleons. Following the holographic prescription, we put the Fock expansion for chiral spinors and the photon bulk-to-boundary propagator into the interaction terms. The hadronic information is encoded in the following invariant functions, whose kernels are the baryonic Schrodinger-like modes and the photon propagator
\begin{eqnarray}
C_1(q)&=&\frac{1}{2}\int{dz\,\left[\psi_L(z)^2+\psi_R(z)^2\right]\,B(z,q)}\\
C_2(q)&=&\frac{1}{2}\int{dz\,e^{A_B(z)}\,\partial_z\,B(z,q)\,\left[\psi_L(z)^2-\psi_R(z)^2\right]}\\
C_3(q)&=&\int{dz\,e^{A_B(z)}\,2\,M_n\,\psi_L(z)\,\psi_R(z)\,B(z,q)}.
\end{eqnarray}
Notice that initial and final baryon states are the same since the collision is elastic. These functions will define the form factors for nucleons as
\begin{equation}
\label{F1_Nuc}
 F_1^N(q)=C_1(q)+\eta_{N}\,C_2(q),\,\,\,\, F_2^N(q)=\eta_{N}\,C_3(q).   
\end{equation}
%
%
From phenomenological arguments, we expect that for protons  $C_1(q)|_{q\to 0}=1$ and $C_2(q)|_{q\to 0}=0$, while for  neutrons $C_1(q)|_{q\to 0}=0$ and $C_2(q)|_{q\to 0}=0$.
  Thus, the electromagnetic form factors for protons are
\begin{equation}
 F_1^p(q)=C_1(q)+\eta_p\,C_2(q),\,\,\,\, F_2^p(q)=\eta_p\,C_3(q).
\end{equation}
%
%
%
By the same token, the electromagnetic form factors for neutrons are defined as
\begin{equation}
F_1^n(q)=\eta_n\,C_2(q),\,\,\,\,   F_2^n(q)=\eta_n\,C_3(q), 
\end{equation}
%
%
Notice that when we evaluate the low $q^2$ limit of $C_3(q)$, we have
\begin{equation}
C_3(0)=2\,M_n\,\int{dz\,e^{A_B(z)}\,\psi_L(z)\,\psi_R(z)}=2\,M_n\,\mathcal{C},    
\end{equation}
where we have used the bulk-to-boundary photon propagator low $q^2$-limit.  In our deformed background model, $\mathcal{C}=2.585$. This expression will be useful when we fix the values of $\eta_{N}$ for protons and neutrons.

Another set of form factors that we can describe are the Sachs electric and magnetic ones, defined for nucleons as
\begin{equation}
G^N_E(q)=F^N_1(q)-\frac{q^2\,}{4\,M_N^2}\,F^N_2(q),\,\,\,\,G^N_M(q)=F^N_1(q)+F^N_2(q),    
\label{SachsFF}
\end{equation}
%
%
which in the low $q^2$ limit define the nucleon magnetic moments as follows
\begin{eqnarray}
G^p_M(0)=\mu_p& &\,\,\text{and}\,\,\quad
G^n_M(0)=\mu_n\,, 
\end{eqnarray}
while for the electric moments
\begin{eqnarray}
G^p_E(0)=1& &\,\,\text{and}\,\,\quad G^n_E(0)=0\,. 
\end{eqnarray}
These conditions allow us to fix $\eta_N$ as
\begin{eqnarray}
\eta_p&=&\frac{1-\mu_p}{2\,M_0\,\mathcal{C}}=0.352\\
\eta_n&=&\frac{\mu_n}{2\,M_0\,\mathcal{C}}=-0.234.
\end{eqnarray}
Note that our model for baryons is invariant under the $SU(2)$ isospin group,  therefore, both proton and neutron have the same mass $M_0$, given by the ground state of the baryonic spectra calculated from Eq. \eqref{scr}.
\begin{center}
\begin{figure}
  \begin{tabular}{c c}
    \includegraphics[width=2.8 in]{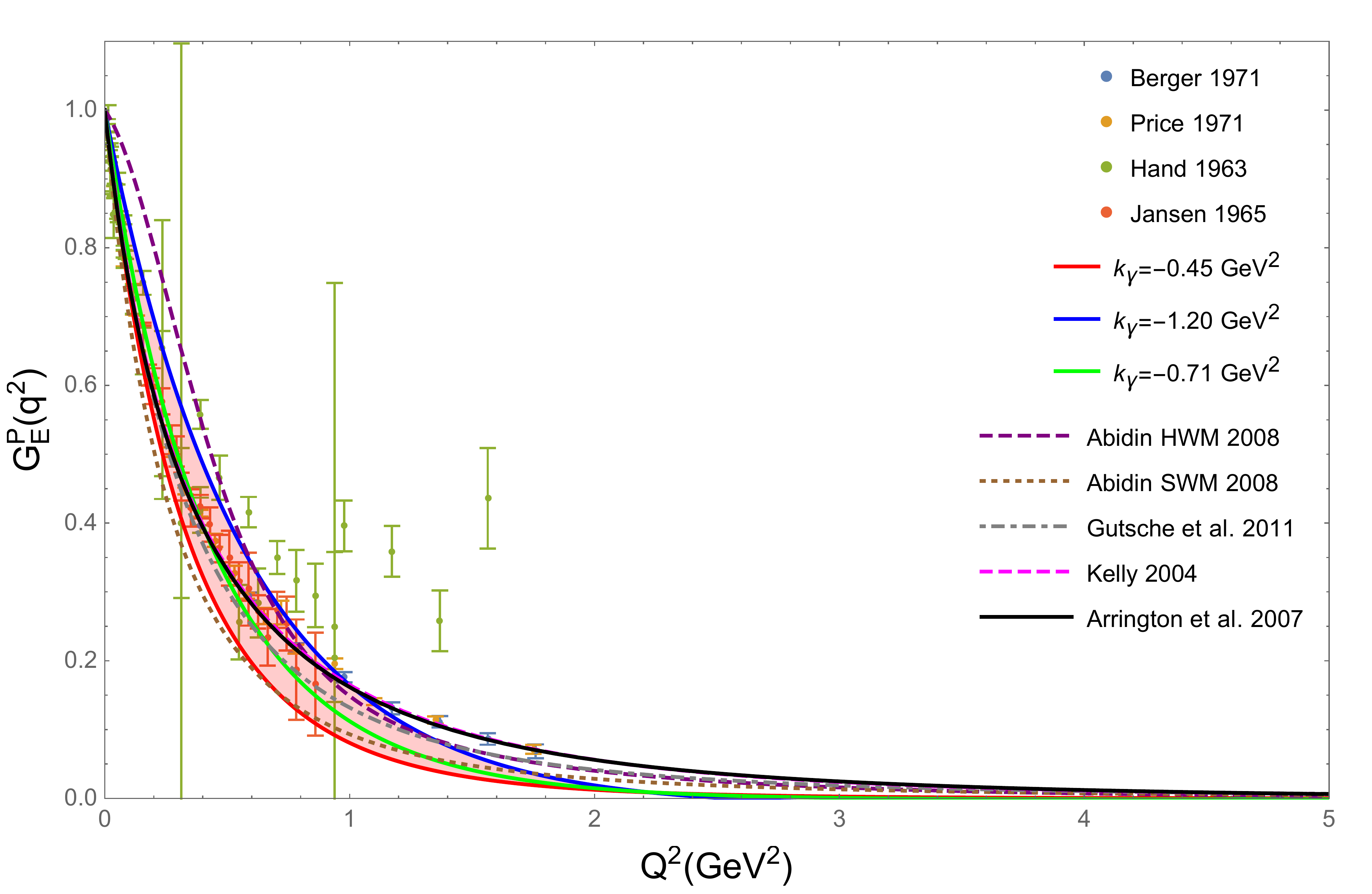}
    \includegraphics[width=2.8 in]{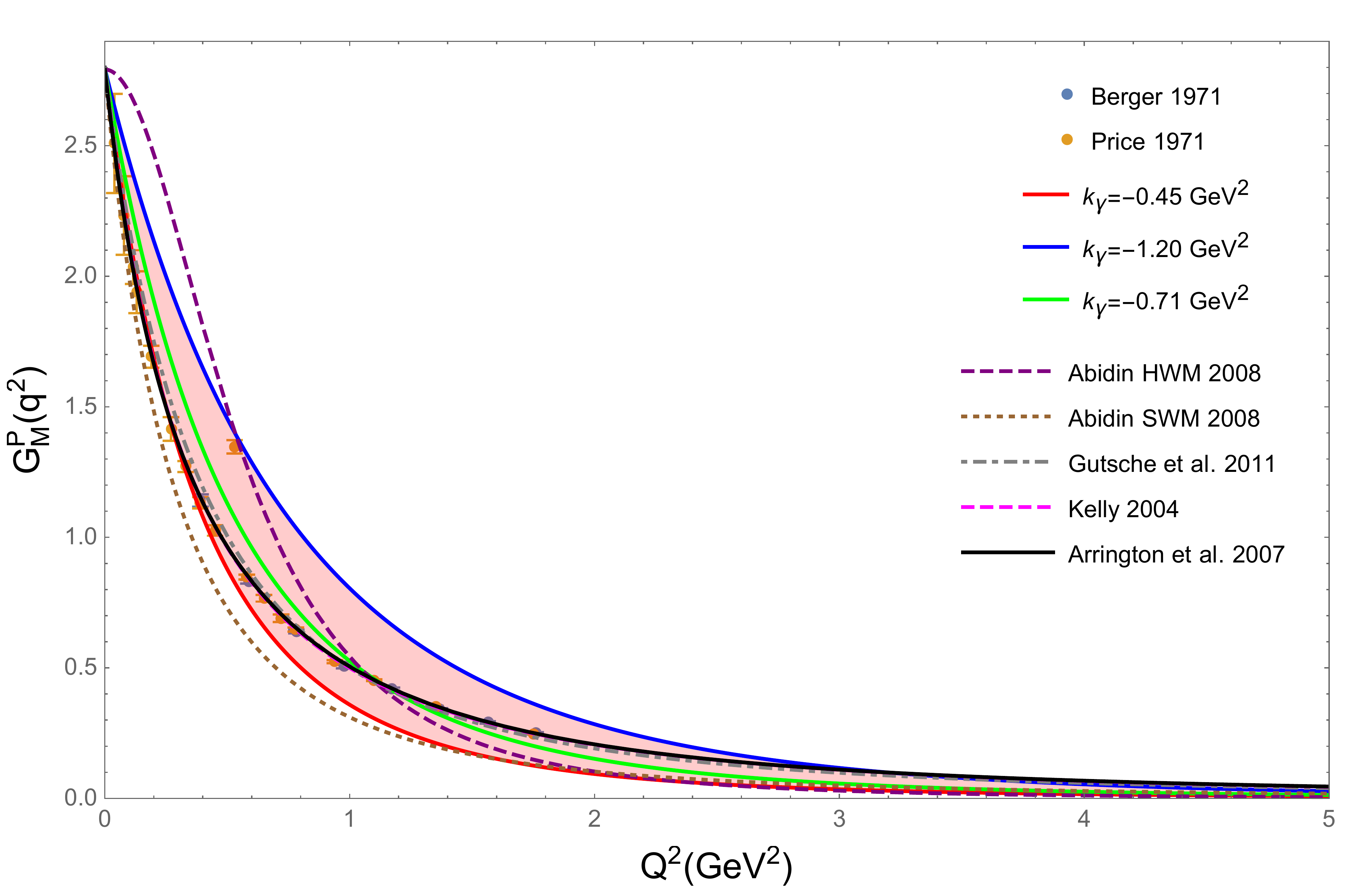}\\
    \includegraphics[width=2.8 in]{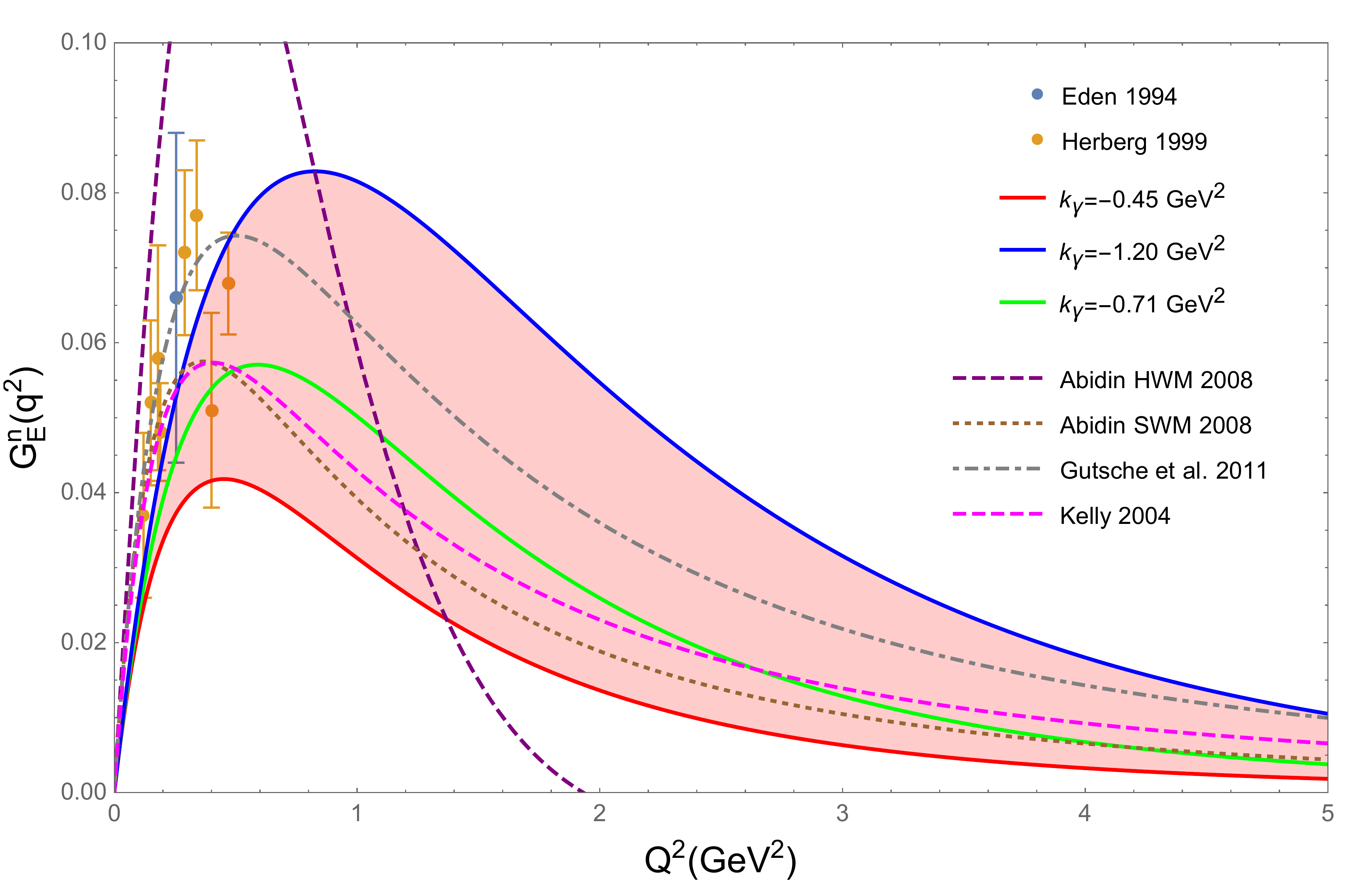}\includegraphics[width=2.8 in]{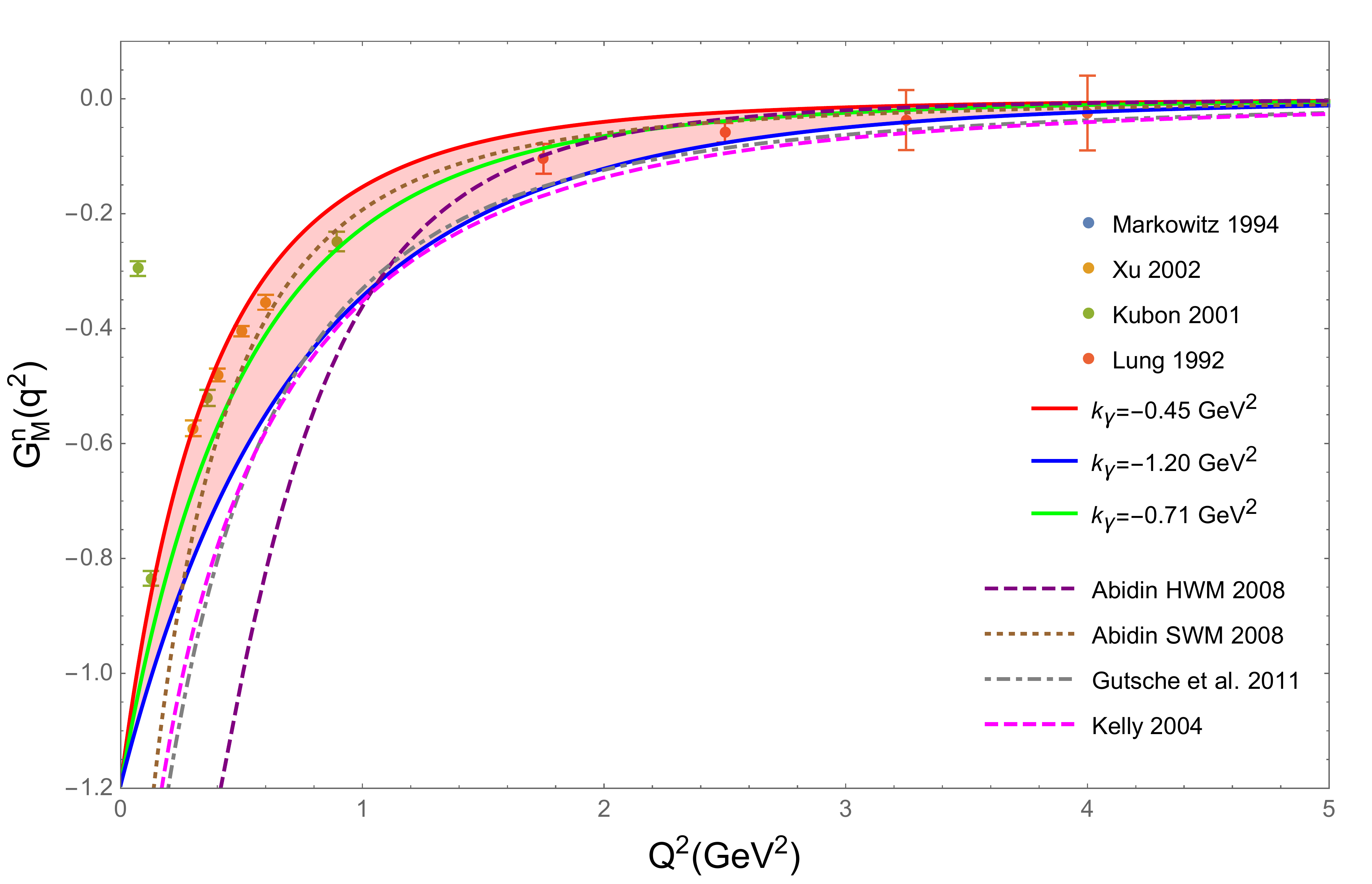}\\

  \end{tabular}
\caption{The upper (lower) panels compare our results for the proton (neutron) Sachs form factors with the available experimental and theoretical data. Left panels: Electric form factors. Right panels: Magnetic form factors.  The experimental data used here are available in Ref.  \cite{Perdrisat:2006hj}.}
\label{fig:one}
\end{figure}
\end{center}

\begin{figure}[!ht]
\begin{center}
\begin{tabular}{c c}
    \includegraphics[scale=0.27]{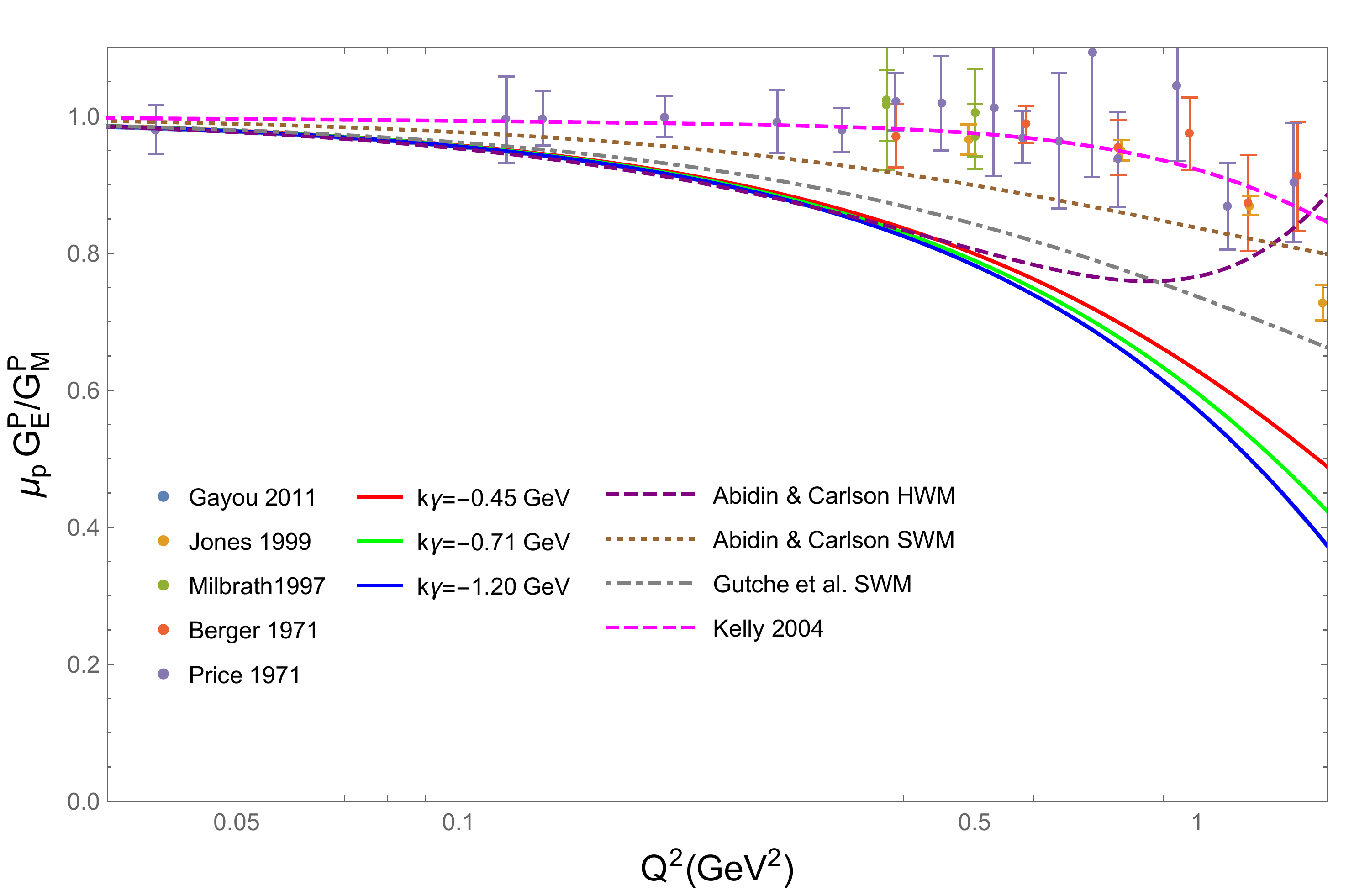} & \includegraphics[scale=0.27]{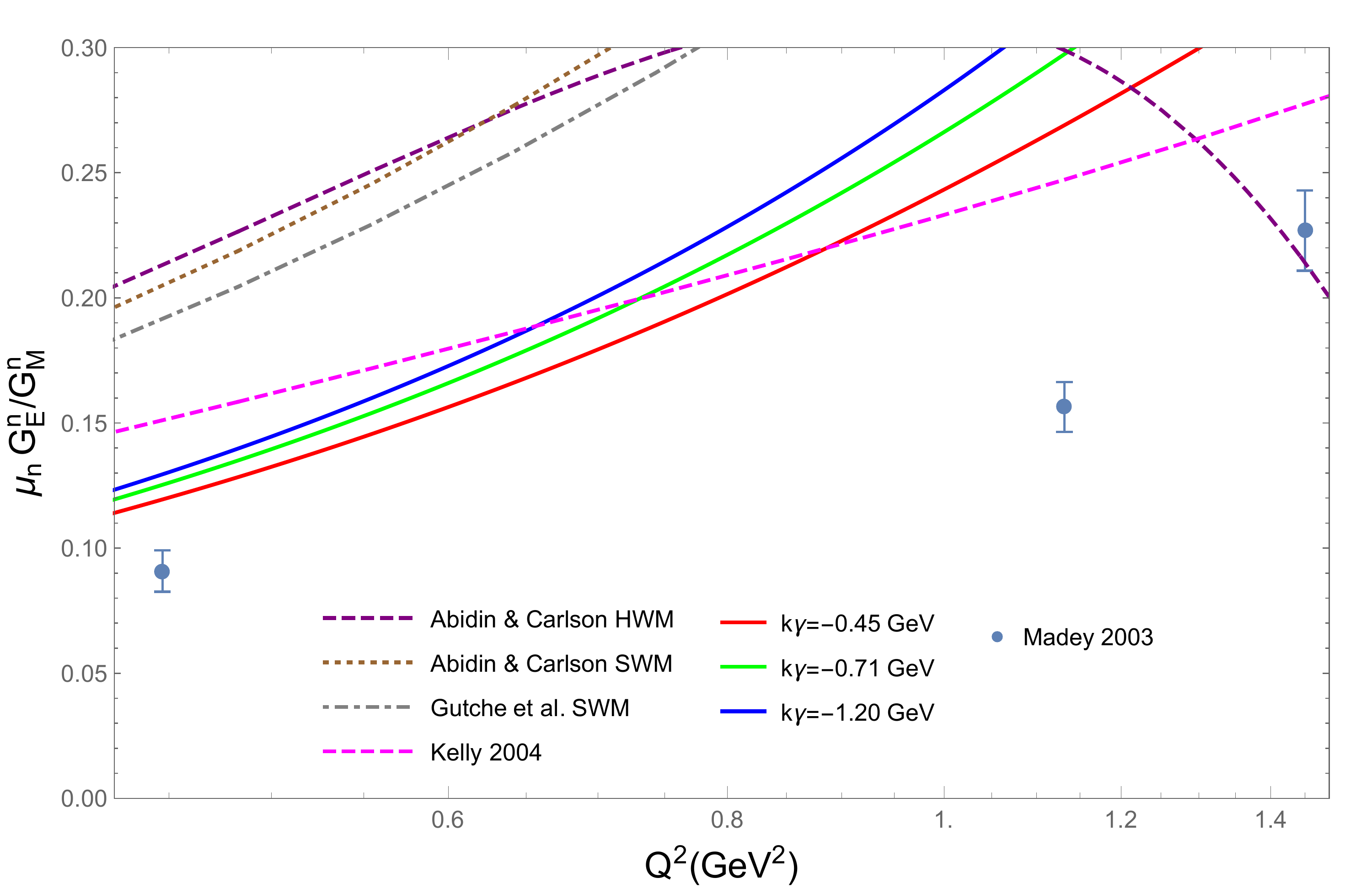} \\
\end{tabular}

\caption{The ratio of the proton (left panel)  and neutron (right panel) Sachs form factors computed within our deformed gravity/gauge duality model  compared with available experimental data and theoretical (holographic) models  \cite{Kelly:2004hm, Perdrisat:2006hj,Abidin:2009hr, Gutsche:2012bp}.}\label{ratio}
\end{center}
\end{figure}

 In Figure \ref{fig:one}, we present our  results for proton and neutron Sachs form factors, Eq. \eqref{SachsFF}, compared with available experimental data and another AdS/QCD model. Note that we have chosen to study a range of values for $k_{\gamma}$ (represented by a colored region) instead of just one value. Of course, one can consider the average value $k_{\gamma} = -0.71$ Gev$^2$ as the best fit. In the left side of Fig. \ref{fig:one}, we show the Sachs electric form factors. In particular, for $G^p_E(q^2)$ (upper left panel) and $G^n_E(q^2)$ (lower left panel) our holographic and deformed model seems to work better for low $q^2$, when it is compared with other models. On the other side, in the upper and lower right panels of Fig. \ref{fig:one}, we plot the Sachs magnetic form factors, where one can see that our deformed model works well from low to large $q^2$.   We compare our results with the ones from  Kelly \cite{Kelly:2004hm}, Arrington et al.  \cite{Arrington:2007ux}, Abidin and Carlson \cite{Abidin:2009hr}, and Gutsche et al.  \cite{Gutsche:2012bp}. We also plot in Fig. \ref{ratio} our predictions for the ratio of the nucleon Sachs form factors and compare them with available experimental data and other  theoretical (holographic)  models. Once again, as shown in Fig. \ref{fig:one}
 our gauge/gravity model as well as another holographic ones seem to work better in the low $q^2$ regime.

Once we have computed the Sachs form factors, we can explore the electromagnetic size of nucleons by computing the corresponding electromagnetic nucleon radii. These quantities, for a given nucleon, are defined as follows
\begin{eqnarray}
\langle r^2_{N,E}\rangle&=&-6\,\left.\frac{d\,G_E^{N}(q^2)}{d\,q^2}\right|_{q^2\to0},\\
\langle r^2_{N,M}\rangle&=&-\frac{6}{G_M^N(0)}\,\left.\frac{d\,G_M^{N}(q^2)}{d\,q^2}\right|_{q^2\to0}.
\end{eqnarray}

Numerical results for the proton and neutron holographic electromagnetic RMS radii are summarized in table \ref{tab:one}.
\begin{center}
\begin{table}[!h]
    \begin{tabular}{||c||c||c||c||}
    \hline
    \hline
    \textbf{Nucleon} & \textbf{Experimental (fm)} & \textbf{This work (fm)}  \\ 
    \hline
    \hline 
       Proton Charge Radius  & $0.8409\pm0.0004$ & $0.8576$  \\ \hline
         Proton Magnetic Radius& $0.851\pm 0.026$ & $0.7929$  \\
         \hline 
         \hline
         Neutron  Charge Radius$^*$ & $-0.1161\pm0.0022$& $-0.0668$ \\\hline
         Neutron Magnetic Radius &$0.864_{-0.008}^{+0.009}$ & $0.7933$ \\
         \hline
         \hline
    \end{tabular}
    \caption{Holographic results are calculated with $\kappa_\gamma=-0.450$ GeV$^2$, in Eq. \eqref{phimunorm}. Experimental data is taken from PDG \cite{Zyla:2020zbs}. For the neutron charge radius, the mean square charge radius, given in fm$^2$, is considered.}
    \label{tab:one}
\end{table}    
\end{center}

\section{Discussions and last comments}\label{results}

In this last section we present our discussions as well as our conclusions on our numerical results for the nucleon form factors and electromagnetic nucleon radii achieved within our model. 
Such a calculation was performed in the context of an AdS/QCD model with a geometric deformation on the metric. Note that in the case of nucleons this deformation sets the confinement scale,while for the photon field, it is related with a kinematical scale of the elastic process. 

First of all, let us discuss about the electric form factors for both protons and neutrons. 
In the low $q^2$ region, holographic models seem to follow a similar phenomenological behavior, despite the fact they are \emph{de facto} different. In Abidin et al. \cite{Abidin:2009hr} and Gutsche et al. \cite{Gutsche:2012bp} and recently in Mamo et al. \cite{Mamo:2021jhj} models, confinement is placed by a quadratic dilaton, whereas our model does the same effect with a geometric (quadratic) deformation. For high $z$, these models have the same asymptotics, as we can see in the linearity of the nucleon spectra \cite{FolcoCapossoli:2019imm}. Differences arise in the intermediate $z$ region, as we can notice by analyzing the e.o.m. structure.  At $z\to0$, these models are entirely dominated by the warp factor only. This condition follows from the consistency constraints we have to address in order to apply the holographic dictionary. However, the intermediate region contributes to set the next to leading order behavior in the form factors at low $q^2$.

Regarding to the Sachs magnetic form factors, in the upper and lower right panels of Fig. \ref{fig:one}, one can see that our deformed model works well for a wide range of $q^2$. 

In Fig. \ref{ratio} we plotted the quotient of $G_E$ with $G_M$ for the proton and the neutron and then we compared them with the experimental data and other holographic and non-holographic approaches.  Notice that for low values of $q^2$, our model had captured the expected behavior of such a quotient. It is worthwhile to mention that this behavior is shared with other holographic bottom-up models. 


In second place, beyond Sachs electric and magnetic form factors another important result achieved in this work is related to the proton and the neutron electromagnetic sizes as summarized in Table \ref{tab:one}. In particular, for proton's charge/magnetic radii and neutron's magnetic radii the results presented here are compatible with the ones in Ref. \cite{Zyla:2020zbs}. Notice again that our model captures the low $q^2$ phenomenology consistently, as our results for the nucleon radii demonstrate. However for the neutron charge radius, our result is not very good. This behavior is also seen in other bottom-up AdS/QCD proposals as presented in Refs.  \cite{Abidin:2009hr,Mamo:2021jhj,Chakrabarti:2013dda}. 
In an overall view, this deformed holographic model presents reasonable results for the electric and magnetic form factors for the proton and the neutron, besides the electric and magnetic nucleon sizes.

\begin{acknowledgments}

A. V. and  M. A. M. C.  would like to thank the financial support given by FONDECYT (Chile) under Grants No. 1180753  and No. 3180592,  respectively. D.L. is supported by the National Natural Science Foundation of China (11805084), the PhD Start-up Fund of Natural Science Foundation of Guangdong Province (2018030310457) and Guangdong Pearl River Talents Plan (2017GC010480). H.B.-F. is partially supported by Coordenação de Aperfeiçoamento de Pessoal de Nível Superior (CAPES) under finance code 001,  and Conselho Nacional de Desenvolvimento Científico e Tecnológico (CNPq) under Grant No. 311079/2019-9.

\end{acknowledgments}

\bibliography{references}
\end{document}